# Atmospheric Retrievals with petitRADTRANS

**Evert Nasedkin**  **[1]¶, Paul Mollière**  **[1], and Doriann Blain**  **[1]**

**1** Max Planck Institut für Astronomie, DE ¶ Corresponding author

## Summary

petitRADTRANS (pRT) is a fast radiative transfer code used for computing emission and transmission spectra of exoplanet atmospheres (Mollière et al., 2019), combining a FORTRAN back end with a Python based user interface. It is widely used in the exoplanet community with 222 references in the literature to date, and has been benchmarked against numerous similar tools, including many listed in (MacDonald & Batalha, 2023). The spectra calculated with pRT can be used as a forward model for fitting spectroscopic data using Monte Carlo techniques, commonly referred to as an atmospheric retrieval (Madhusudhan & Seager, 2009). The new retrieval module combines fast forward modelling with nested sampling codes, allowing for atmospheric retrievals on a large range of different types of exoplanet data. Thus it is now possible to use pRT to easily and quickly infer the atmospheric properties of exoplanets in both transmission and thermal emission.

## Statement of need

Atmospheric retrievals are a cornerstone of exoplanet atmospheric characterisation. Previously this required interfacing with a sampler and writing a likelihood function in order to explore a parameter space and fit input data, following the description provided in (Mollière et al., 2019). pRT now provides a powerful and user-friendly tool for researchers to fit exoplanet spectra with a range of built-in or custom atmospheric models. Various thermal structures, chemistry and cloud parameterisations and opacity calculation methods can be combined and used to perform parameter estimation and model comparison for a given atmospheric spectrum. The new retrieval module standardises data input, parameter setup, likelihood calculation and interfaces with Multinest, allowing for both simple 'out-of-the-box' retrievals as well as fully customisable setups. New tutorials and documentation have been published to facilitate the adoption of the retrieval module as a widely used tool in the exoplanet community. With increasing volumes of both ground- and space-based spectra available, it is necessary for exoplanet researchers to have access to a range of characterisation tools. While many other retrieval codes exist, pRT offers a computationally efficient and highly flexible framework for inferring the properties of a wide variety of exoplanets. While most feature of pRT are available in other retrieval codes, few of the implement the diversity of the pRT feature set; the availability of both correlated-k and line-by-line opacities, free, equilibrium, and disequilibrium chemistry, and multiple cloud implementations in a single framework let it stand out as a uniquely flexible approach to atmospheric retrievals.

## petitRADTANS Retrieval Module

The new retrieval module combines the existing Radtrans forward modelling class with a nested sampler via a likelihood function to perform an atmospheric retrieval. Both MultiNest (Buchner et al., 2014; Feroz et al., 2009, 2019; Feroz & Hobson, 2008) and Ultranest (Buchner et al., 2014; Buchner, 2019) samplers are available, with both offering MPI implementations that allow for easy parallelisation.

Datasets, priors and other retrieval hyper parameters are set through the `RetrievalConfig` class, while the `models` module includes a range of complete atmospheric models that can be fit to the data. Users can also define their own model function, either by making use of temperature profiles from the `physics` module and chemistry parameterisations from the `chemistry` module or by implementing their own forward model. Once set up, the `Retrieval` class is used to run the retrieval using the chosen sampler and to produce publication ready figures.

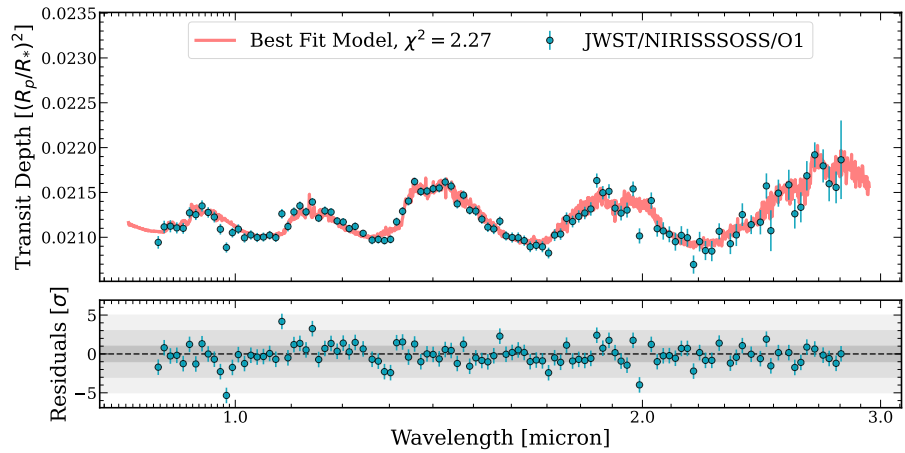

**Figure 1:** Typical example of default pRT outputs. This highlights the fit of a transmission spectrum model to JWST/NIRISS/SOSS data of WASP 39 b as part of the Transiting Early Release Science program.

Multiple datasets can be included into a single retrieval, with each dataset receiving its own `Radtrans` object used for the radiative transfer calculation where some or all forward model parameters may be shared between the different data sets. This allows for highly flexible retrievals where multiple spectral resolutions, wavelength ranges and even atmospheric models can be combined in a single retrieval. Each dataset can also receive scaling factors (for the flux, uncertainties or both), error inflation factors and offsets. The model functions are used to compute a spectrum $\vec{S}$, which is convolved to the instrumental resolution and binned to the wavelength bins of the data using a custom binning function to account for non-uniform bin sizes. The resulting spectrum compared to the data with flux $\vec{F}$ and covariance $\mathbf{C}$ in the likelihood function:

$$-2\log \mathcal{L} = \left(\vec{S} - \vec{F}\right)^T \mathbf{C}^{-1} \left(\vec{S} - \vec{F}\right) + \log\left(2\pi \det\left(\mathbf{C}\right)\right). \qquad (1)$$

The second term is included in the likelihood to allow for uncertainties to vary as a free parameter during the retrieval, and penalises overly large uncertainties. A likelihood function for high resolution data based on that of (Gibson et al., 2020) is also available.

pRT can compute spectra either using line-by-line calculations, or using correlated-k (c-k) tables for defining the opacities of molecular species. We include up-to-date correlated-k line lists from Exomol (Chubb et al., 2021; McKemmish et al., 2016; Polyansky et al., 2018; Tennyson & Yurchenko, 2012) and HITEMP Hargreaves et al. (2020), with the full set of available opacities listed in the online documentation. The exo-k package is used to resample the correlated-k opacity tables to a lower spectral resolution in order to reduce the computation time (Leconte, 2021). Combining the c-k opacities of multiple species requires mixing the distributions in $g$ space. This operation is necessary when calculating emission spectra and accounting for multiple scattering in the clouds. Previously, this was accomplished by taking 1000 samples of each distribution. This sampling process resulted in non-deterministic spectral calculations with a small (up to 1%) scatter about the expected result, which could lead to unexpected behaviour from the nested sampler as the same set of parameters could result in different log-likelihood. pRT has been updated to fully mix the c-k distributions, iteratively mixing in any species with a significant opacity contribution: a species is only mixed in if the highest opacity value in a given spectral bin is larger than a threshold value. This threshold value is obtained by listing the smallest opacity value for every species in a given spectral bin, and then setting the threshold to 1% of the largest value from the list for each spectral bin. The resulting grid is linearly interpolated back to the 16 $g$ points at each pressure and frequency bin as required by pRT. This fully deterministic process produces stable log-likelihood calculations, and resulted in a 5$\times$ improvement in the speed of the c-k mixing function.

Various thermal, chemical and cloud parameterisations are available in pRT. Built-in temperature profiles range from interpolated splines to physically motivated profiles as in Guillot ([2010](#)) and Mollière et al. ([2020](#)). Equilibrium and disequilibrium chemistry can be interpolated from a pre-computed grid on-the-fly. Chemical abundances can also be freely retrieved, with the additional possibility of using a combination of free and chemically consistent abundances. Cloud parameterisations range from a 'grey' continuum opacity applied at all wavelengths, to clouds parameterised as in Ackerman & Marley ([2001](#)), using log-normal or Hansen ([1971](#)) particle size distributions with real optical opacities for different compositions and particle shapes, and including self-scattering. Users can also pass functions to the Radtrans object that encode any absorption or scattering opacity as a function of wavelength and atmospheric position, allowing for a generic cloud implementation. Included in pRT is an option to use an adaptive pressure grid with a higher resolution around the location of the cloud base, allowing for more precise positioning of the cloud layers within the atmosphere.

Photometric data are fully incorporated into the retrieval process. The spectral model is multiplied by a filter transmission profile from the SVO database using the `species` package ([Stolker et al., 2020](#)). This results in accurate synthetic photometry, which can be compared to the values specified by the user with the `add_photometry` function.

Publication-ready summary plots of best fits, temperature and abundance profiles and corner plots can be automatically generated. Multiple retrieval results can be combined in the plots for model comparisons. Such results have been benchmarked against other widely used retrieval codes, such as `PLATON` ([Zhang et al., 2019](#)), `POSEIDON` ([Grant et al., 2023](#)) and `ARCiS` in ([Dyrek et al., 2024](#)). The forthcoming retrieval comparison of the JWST Early Release Science (ERS) program will comprehensively compare pRT and other retrieval codes in the analysis of WASP-39b (Welbanks et al., in prep). Figure [1](#) shows the fit of a transmission model to the JWST/NIRISS/SOSS observations of WASP 39 b ([Feinstein et al., 2023](#)) from the ERS program.

## Acknowledgements

We gratefully acknowledge contributions to petitRADTRANS from Eleonara Alei, Karan Molaverdikhani, Francois Rozet, Aaron David Schneider, Tomas Stolker, Nick Wogan and Mantas Zilinskas. Thanks as well to our reviewers, Quentin Changeat and Valentin Christiaens, as well as our editor Dan Foreman-Mackey.